\documentclass[12pt]{iopart}
\usepackage{iopams}
\usepackage{graphicx}
\usepackage{color}
\usepackage{times}
\usepackage[colorlinks,bookmarks=false,citecolor=blue,linkcolor=red,
urlcolor=blue]{hyperref}

\usepackage{float}

\newcommand{\perpp}{\scriptscriptstyle{\perp}}
\newcommand{\parallell}{\scriptscriptstyle{\parallel}}
\newcommand{\Jperp}{J_{\perpp}}

\newcommand{\jC}{j_{\textrm{\tiny C}}}
\newcommand{\jR}{j_{\textrm{\tiny R}}}

\begin{document}

\title{Strongly interacting bosons on a three-leg ladder in the presence of a homogeneous flux}

\author{F Kolley$^1$, M Piraud$^1$\footnote[1]{To
whom correspondence should be addressed (marie.piraud@physik.uni-muenchen.de)},
I P McCulloch$^2$, U Schollw\"ock$^1$, F Heidrich-Meisner$^1$}
\address{$^1$Department of Physics and Arnold Sommerfeld Center 
for Theoretical Physics, Ludwig-Maximilians-Universit\"at M\"unchen, 
D-80333 M\"unchen, Germany}
\address{$^2$ARC Centre for Engineered Quantum Systems, School of Mathematics and Physics, The University of Queensland, St Lucia, Queensland 4072, Australia}

\begin{abstract}
We perform a density-matrix renormalization-group study of strongly interacting bosons on a three-leg ladder in the presence of a homogeneous flux.
Focusing on one-third filling, we explore the phase diagram in dependence of the magnetic flux and the inter-leg tunneling strength.
We find several phases including a Meissner phase, vortex liquids, a vortex lattice, as well as a staggered-current phase.
Moreover, there are regions where the chiral current  reverses its direction, both in the Meissner and in the staggered-current phase.
While the reversal in the latter case can be ascribed to spontaneous breaking of  translational invariance, in the first it stems from an effective flux increase in the rung direction.
Interactions are a necessary ingredient to realize either type of chiral-current reversal.
\end{abstract}

\maketitle

\section{Introduction}
Experimental progress with ultracold quantum gases has made feasible  engineering the coupling between the different states of the atoms, in order to realize synthetic gauge fields~\cite{lin09,dalibard11}.
The effective magnetic fields acting on  the neutral atoms can be  much larger than what is possible in solid-state systems.
These advances bring the simulation of a wide range of Hamiltonians into reach that are important in condensed matter physics~\cite{lin11, aidelsburger11, jimenez12, struck12, aidelsburger13, miyake13}.
Indeed, some of the most intriguing phenomena in condensed matter physics involve the presence of strong magnetic fields.
For instance, topological  
states of matter are realized in quantum Hall systems~\cite{thouless82,hasan10},
which are insulating in the bulk, but bear conducting edge states. 
Remarkably, topological phase transitions were observed in experiments with cold atoms~\cite{atala13, jotzu14, aidelsburger15}.

Recently, there has been a growing theoretical (see, e.g., work on fermions~\cite{carr06,roux07,sun13} and bosons~\cite{dhar12,dhar13,petrescu13,grusdt14b,piraud15,greschner15,didio15}) and experimental~\cite{atala14,mancini15,stuhl15} interest in quasi-one dimensional relatives of  the square lattice~\cite{harper55,hofstadter76,jaksch03,sorensen05,palmer06,palmer08,hafezi07,moeller09,senthil13,baur13,grusdt14a}, 
the $N$-leg ladder systems with synthetic gauge fields.
Besides using superlattices~\cite{atala14} to realize these geometries, 
 a synthetic lattice dimension can be exploited where the sites on the rungs of the ladder correspond to different hyperfine states~\cite{celi14}.
The latter has motivated several recent theoretical studies~\cite{zeng15,yan15,ghosh15,qin15,barbarino15,natu15}.
Chiral edge currents on two- and three-leg ladders subjected to a homogeneous flux have been observed using both experimental approaches, for bosons \cite{atala14,stuhl15} and fermions \cite{mancini15}.
The physics of interacting bosons on ladders has also been studied in the context of Josephson junction arrays~\cite{kardar86,mazo95,trias2000,binder2000}, albeit in the weakly-interacting regime.

Such $N$-leg ladder systems, similar to their counterparts in quantum magnetism~\cite{dagotto99}, provide an elegant bridge between the physics in one and two dimensions~\cite{dagotto95}.
Interacting bosons on two-leg ladders in homogeneous magnetic fields harbor very rich physics,  featuring  Meissner and vortex phases~\cite{orignac01,dhar12, dhar13,petrescu13,huegel14,tokuno14,piraud15,didio15}, a biased-ladder phase that breaks the symmetry between the two legs~\cite{wei14,uchino15,greschner15}, analogues of Laughlin states~\cite{petrescu15} as well as an interaction-driven reversal of the direction of the current due to spontaneous breaking of translational invariance in vortex lattices~\cite{greschner15}.

Here, we use density-matrix renormalization-group (DMRG)~\cite{white92, schollwoeck05, schollwoeck11} simulations to explore the phase diagram of strongly interacting bosons on a three-leg ladder subjected to a homogeneous flux.
We focus on a filling of one-third of a boson per site, which can easily be realized in experiments~\cite{stuhl15}.
At this filling, the system is expected to be in a Mott-insulating state, based on the existence of a magnetization plateau at one-third
of the saturation magnetization in the closely related  three-leg spin-1/2 ladders~\cite{cabra97,cabra98} and on work on the MI-SF transition in bosonic three-ladders at zero flux a zero flux~\cite{ying14}.
We present the  phase diagram for hard-core bosons (HCBs) in dependence of the inter-leg coupling strength and the magnitude of the magnetic flux.
As a main result, we observe vortex phases that cannot be traced back to features in the single-particle dispersion.
In the Meissner phase in this system, 
there is a fascinating reversal of the direction of the current, driven by  the magnetic flux.
We provide an explanation for this effect and we show that it  persists at intermediate interaction strengths.
Finally, a staggered-current phase, generalizing the one found on the two-leg ladder~\cite{dhar12, dhar13}, is observed around $\phi \sim \pi$,
which also triggers a chiral-current reversal.
%
\begin{figure}[t]
\centering
\includegraphics[width=0.7\linewidth]{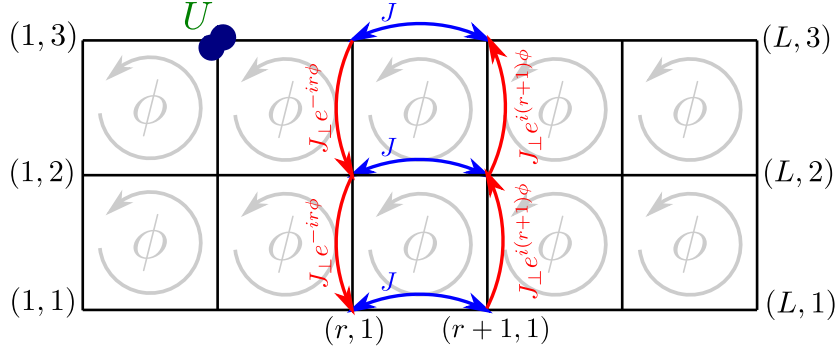}
\caption{
Schematic representation of the three-leg ladder model~(\ref{eq:hamiltonian}).
$(r,\ell)$ labels the site on the $r$-th rung and on the $\ell$-th leg.
The tunneling strength along the legs(rungs) is $J$($\Jperp$).
We choose the gauge such that  the phase is picked up on the rungs, resulting in a net flux of $\phi$ through each plaquette.
The on-site interaction strength is given by $U$.
}
\label{fig:lattice}
\end{figure}
%

\section{Model and method}
We consider the Bose-Hubbard Hamiltonian on  a three-leg ladder of length $L$ (see figure~\ref{fig:lattice}):
\begin{eqnarray}\label{eq:hamiltonian}
&\mathcal H =-\Jperp\sum_{r=1}^{L}\sum_{\ell=1}^{2}\left( e^{-ir\phi}a^{\dagger}_{r,\ell}a_{r,\ell+1} + h.c.\right) \\
& -J \sum_{r=1}^{L-1}\sum_{\ell=1}^{3} \left(a^{\dagger}_{r,\ell}a_{r+1,\ell} + h.c. \right) + \frac{U}{2} \sum_{r=1}^{L}\sum_{\ell=1}^{3} n_{r,\ell}(n_{r,\ell}-1) \, , \nonumber
\end{eqnarray}
where $a^{\dagger}_{r,\ell}$ creates a boson on the $r$-th rung and the $\ell$-th leg of the system and $n_{r,\ell}\equiv a^{\dagger}_{r,\ell}a_{r,\ell}$ is the on-site number operator.
$J$ and $\Jperp$ are the tunneling matrix elements  along the legs and rungs, respectively,
 $U$ is the on-site interaction strength, and $\phi$ is the magnetic flux through a plaquette of the ladder.

We compute the ground state of~\eref{eq:hamiltonian} numerically with a single-site DMRG algorithm~\cite{white2005}. While exploiting the $U(1)$ symmetry of the Hamiltonian associated to particle number conservation, we keep up to $4000$ DMRG states and
the data in the figures of the main text are for $L=100$, while we have considered $L$ as large as $L=200$ in the appendix.
For $U/J < \infty$, we limit the number of bosons per site to  six at maximum. For the determination of the Mott insulator (MI) to superfluid (SF) transitions, we use infinite-size DMRG~\cite{mcculloch08} with up to 600 DMRG states [see \ref{sec:appendix_massgap}].

The different phases are primarily characterized by their local current configurations.
The associated operators are
\begin{eqnarray}
j^{\parallell}_{r,\ell} &= iJ\left( a^{\dagger}_{r+1,\ell}a_{r,\ell} - a^{\dagger}_{r,\ell}a_{r+1,\ell} \right)\, ,\\
j^{\perpp}_{r,\ell} &= i\Jperp\left(e^{-ir\phi} a^{\dagger}_{r,\ell+1}a_{r,\ell} - e^{ir\phi}a^{\dagger}_{r,\ell}a_{r,\ell+1} \right)\, .
\end{eqnarray}
We  compute the chiral current from 
\begin{eqnarray}\label{eq:meissner}
 \jC = \frac{1}{2L} \sum_{r}\langle j^{\parallell}_{r,1} - j^{\parallell}_{r,3}\rangle .
\end{eqnarray}
Throughout the paper, we fix   
the  filling to $1/3$ bosons per site.

%
\begin{figure}[t]
\centering
\includegraphics[width=.7\linewidth]{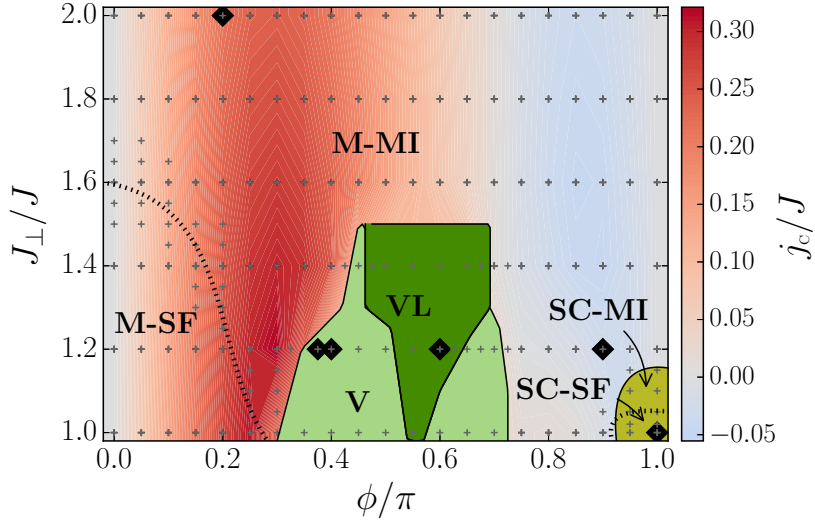}
\caption{Phase diagram for HCBs with $1/3$ particle per site on average, as a function of $\Jperp$ and $\phi$.
The dotted lines are upper bounds for the Mott-insulator (MI) to superfluid (SF) transition, as determined by an analysis of the Binder cumulant corresponding to an appropriate string-order parameter [see \ref{sec:appendix_massgap}]. The phase diagram features Meissner (M-MI/M-SF), vortex-liquid (V), vortex-lattice (VL) and staggered-current (SC-MI/SC-SF) phases.
While the SC region is split into a Mott insulating and a superfluid phase, the V and VL phases are Mott insulating,
In the Meissner phase, the color encodes the amplitude of the chiral current $\jC$.
The gray '+' symbols mark the parameter points analyzed with DMRG.
Large black diamonds denote the parameter values for which the local-current patterns are presented in figure~\ref{fig:currentpatterns}.
The vortex-liquid region contains one- and two-component Luttinger liquids.
The VL is expected to be surrounded by a (possibly very thin) vortex-liquid phase.
}
\label{fig:phasediagram}
\end{figure}
%

\section{Phase diagram}
\subsection{Summary of results for two-leg ladders}

Let us first briefly summarize some results for the corresponding two-leg ladder model in the low density regime (obtained by limiting the number of legs to two in figure~\ref{fig:lattice}), which will help to appreciate our results for the three-leg ladder to be
presented in the following.
The phase diagram for hard-core bosons at a filling of one boson per rung has been reported in~\cite{piraud15,didio15}: at fixed rung coupling $\Jperp/J <1.6$, there is a transition from a phase with Meissner-like currents to a vortex phase when increasing the flux.
Above a critical value ${\Jperp}^{\rm c} \simeq 1.6 J$, the Meissner phase is stable at any flux.
Furthermore, at certain commensurate vortex densities, vortex lattices are expected to form~\cite{orignac01}, which, however, do
not seem to exist in the limit of   hard-core bosons \cite{piraud15}. At smaller interaction strength, though, 
stable vortex lattices at vortex densities of $1/2$ and $1/3$ have been identified in~\cite{greschner15}. 
In  those phases, the translation invariance associated with translations along the legs of the ladder is spontaneously broken and an enlarged unit cell forms.
As a consequence,  the effective flux sensed by the bosons is modified. This leads to a spontaneous reversal of  the direction of the chiral current, under certain conditions on the flux and interaction strength~\cite{greschner15}.

\subsection{Overview of the phase diagram for the three-leg ladder}
Figure~\ref{fig:phasediagram} shows the ground-state phase diagram for HCBs [$(a^{\dagger}_{r,\ell})^2=0 $;  $U/J \to \infty$] of the three-leg ladder model equation~\eref{eq:hamiltonian} as a function of $\Jperp \in [J,2J]$ and flux.
Results are displayed for $\phi \in [0,\pi]$ since all physical quantities are $2\pi$-periodic, odd or even, functions of the flux~\cite{greschner15}.
We find the following phases: (i) a Meissner phase (M-MI/M-SF),
which shows a reversal of the current direction for large values of the flux,
(ii) vortex-liquid phases (V),
(iii) a vortex-lattice phase (VL),
and (iv) staggered-current phases (SC-MI/SC-SF).
The corresponding transitions are located by solid lines in Fig.~\ref{fig:phasediagram}, in which the gray '+' symbols indicate the parameter points that were analyzed numerically.
For $\Jperp/J \gtrsim 1.6$ or $\phi \gtrsim 0.3 \pi$, we observe a finite mass gap [see \ref{sec:appendix_massgap}] indicating that the system is in a Mott-insulating state.
This is inherited from  the $\Jperp/J \gg 1$ limit, in which the ground state is the product  of the local ground state on each rung, separated by a gap $\sqrt{2} \Jperp$ from all excited states (similar to the two-leg ladder with one boson per rung~\cite{crepin11}).
Upon lowering $\Jperp$ at fixed $\phi$, the mass gap eventually closes.
The computation of the Binder cumulant~\cite{binder81} of an appropriate string-order parameter for the Mott insulator yields an upper bound for the transition to a superfluid phase when lowering $\Jperp$ (see the dotted line in figure~\ref{fig:phasediagram}).
This transition is, for $\phi/\pi \lesssim 0.3$, compatible with a Berezinskii-Kosterlitz-Thouless transition [see \ref{sec:appendix_massgap}] as expected from the theoretical work on magnetization plateaux in $N$-leg spin ladders~\cite{cabra97,cabra98} and numerical work at $\phi=0$~\cite{ying14}.
The transition at $0.9 < \phi/\pi \leq 1$, on the other hand, is compatible with a second-order phase transition [see \ref{sec:appendix_massgap}].
In the following, we refer to states with a Meissner-like  current configurations using the label M, and to states with a staggered-current pattern
using the label SC (irrespective of whether these are MI or SF phases).
%
\begin{figure}[t]
\centering
\includegraphics[width=1.\linewidth]{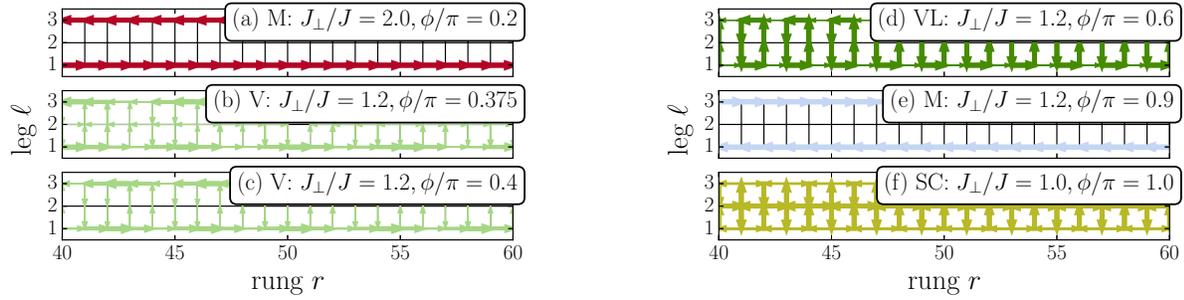}
\caption{Current patterns for HCBs in
 (a) the Meissner phase ($\Jperp/J=2$, $\phi/\pi=0.2$); (b-c) in the  vortex-liquid phases ($\Jperp/J=1.2$, $\phi/\pi=0.375$ and $0.4$); (d) in the VL phase ($\Jperp/J=1.2$, $\phi/\pi=0.6$); 
(e) in the Meissner phase, where the chiral current is reversed  ($\Jperp/J=1.2$, $\phi/\pi=0.9$); and  (f) in the staggered-current phase ($\Jperp/J=1$, $\phi=\pi$).
 The width of the arrows is proportional to the current strength and normalized to the strongest current in the region displayed for each parameter point. 
}
\label{fig:currentpatterns}
\end{figure}
%
\subsection{Meissner phases}
For $\Jperp/J \geq 1.6$, only a Meissner phase exists, characterized by currents occurring exclusively in the upper and the lower leg [see figures~\ref{fig:currentpatterns}(a) and~\ref{fig:currentpatterns}(e)], with a finite mass gap.
The existence of such a Meissner-Mott insulator (M-MI)  on the three-leg ladder has been predicted in reference~\cite{petrescu15}. For $\phi/\pi \leq 0.3$, we find a transition from this Mott-insulator to a superfluid phase with Meissner currents (M-SF) by lowering the inter-chain coupling $\Jperp$ [dashed line in \fref{fig:phasediagram}].
Intriguingly, the chiral current reverses its chirality from counterclockwise [figure~\ref{fig:currentpatterns}(a)] to clockwise [figure~\ref{fig:currentpatterns}(e)] for $\phi/\pi \gtrsim 0.75$, 
meaning that the atoms  flow in the direction opposite to the one favored by the effective magnetic field.
As a consequence, at the  point at which the reversal occurs, no current flows  even though the bosons feel a very strong, non-staggered, magnetic flux.
A typical example for the $\jC=\jC({\phi})$ curve in the Meissner phase is shown in figure~\ref{fig:meissner}(a) for $\Jperp/J=1.8$. The curve is smooth as there is no phase transition.

\subsection{Vortex phases}
\begin{figure}[t]
\centering
\includegraphics[width=.7\linewidth]{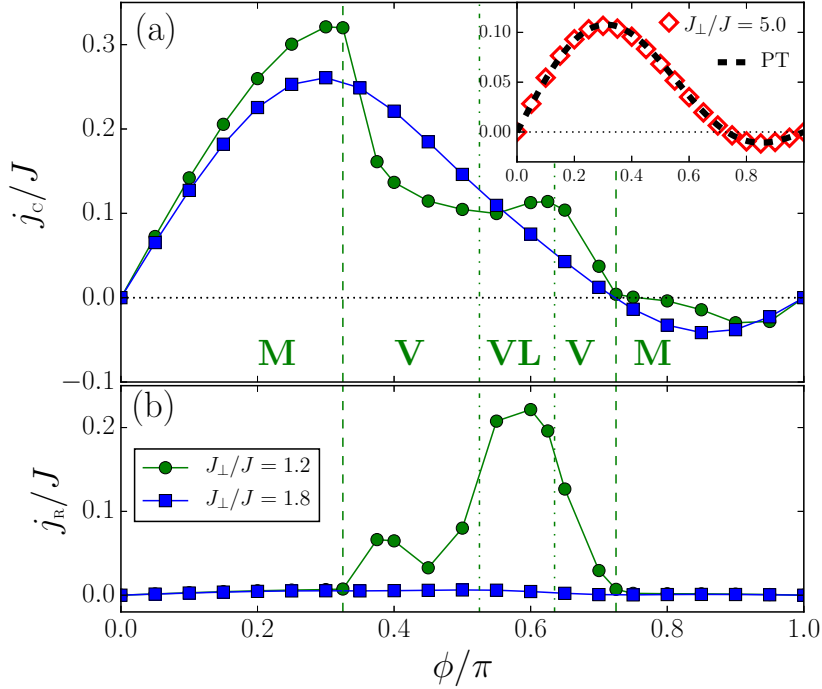}
\caption{(a) Chiral current $\jC$ [cf.~equation~\ref{eq:meissner}] as a function of the magnetic flux $\phi$ for $\Jperp/J=1.2$ and $1.8$ and HCBs. Inset: 
Comparison with perturbation theory (PT) [equation~\ref{eq:PT}] for $\Jperp/J = 5$. (b) Average rung current $\jR=\sum_{r,\ell=1,2} | \langle j^{\perpp}_{r,\ell} \rangle | /2L$ as a function of flux $\phi$ for $\Jperp/J=1.2$ and $1.8$.
}
\label{fig:meissner}
\end{figure}

Upon lowering $\Jperp$ to $\Jperp/J \lesssim 1.5 $,  the Meissner phase is split by vortex phases at intermediate values of the flux.
Typical  current patterns in the vortex phases, which, for open boundary conditions, exhibit 
non-zero rung currents $\langle j^{\perpp}_{r,\ell} \rangle$,   are presented in figures~\ref{fig:currentpatterns}(b)-(d).
We also plot the chiral current $\jC$ [see figure~\ref{fig:meissner}(a)] as a function of $\phi$ for $\Jperp/J= 1.2$, together with the average absolute value of the rung current $\jR=\sum_{r,\ell=1,2} | \langle j^{\perpp}_{r,\ell} \rangle | /2L$ [see figure~\ref{fig:meissner}(b)].
The transition from the Meissner into the vortex phases is characterized by  $\jR$ becoming non-zero and by  kinks in the $\jC=\jC(\phi)$ curves (see the example of $\Jperp/J=1.2$).
%
%
The vortex phases can be further divided into vortex-liquid (V) phases, which are incommensurate, gapless phases [see, e.g., figure~\ref{fig:currentpatterns}(b)-(c)] and a vortex lattice (VL), which is fully gapped and forms at a commensurate vortex density~\cite{orignac01} [see, e.g., figure~\ref{fig:currentpatterns}(d)].

\subsubsection{Vortex lattice.}
The commensurability of the phases can be unveiled by  studying the spatial patterns of the rung currents $\langle j^{\perpp}_{r,\ell=1} \rangle$.
In general,  their Fourier transforms bear two typical wavelengths $q_A$ and $q_B$ (see the inset of figure~\ref{fig:vortex-dens}, we  choose $|q_B| < |q_A|$). 
Motivated by this observation, we define two vortex densities $\rho_v^{A,B}=q_{A,B}a^{-1}/(2\pi)$ (with $a$ the lattice spacing), which are shown in figure~\ref{fig:vortex-dens} for $\Jperp/J=1.2$.
In the Meissner and VL phases, both vortex densities are commensurate with  $\rho_v^{\alpha}=0$ and $\rho_v^{\alpha}=1/2$ (with $\alpha=A, B$), respectively. 
These two phases are fully gapped. In the case of the VL, we ascribe this behavior to the   incompressibility of the vortex pattern: 
it costs a finite energy to add a vortex to the system, even in the thermodynamic limit.
The transitions to the vortex-liquid phases are  continuous commensurate-incommensurate transitions~\cite{orignac01}.
We therefore expect the vortex-liquid region to surround the VL phase everywhere, even though the proximity of various phase transitions renders it very difficult to resolve numerically.

\subsubsection{Vortex liquids.}
\begin{figure}[t]
\centering
\includegraphics[width=.7\linewidth]{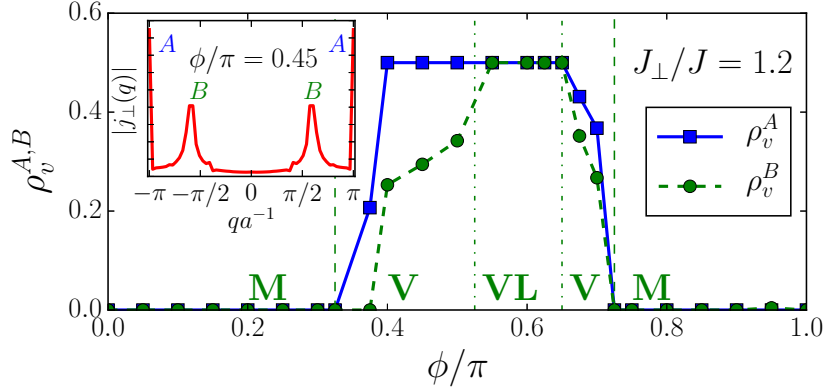}
\caption{Vortex densities $\rho^{A,B}_{v}$ as a function of magnetic flux $\phi$ for $\Jperp/J=1.2$ and HCBs. They are extracted from the Fourier transform of $\langle j^{\perpp}_{r,\ell=1} \rangle$, which show two dominant wavelengths in the vortex phase as illustrated in the inset for $\phi/\pi=0.45$.
}
\label{fig:vortex-dens}
\end{figure}

The incommensurate vortex-liquid region (V) encompasses phases in which both vortex densities are incommensurate [e.g., at $\phi/\pi=0.7$ in figure~\ref{fig:vortex-dens}]
as well as phases where one mode is commensurate (at either $\rho_v^{\alpha}=0$ or $\rho_v^{\alpha}=1/2)$ and the other is not [e.g., at $\phi/\pi=0.375$ and $0.5$ in figure~\ref{fig:vortex-dens}].
This is fully corroborated by the study of the von Neumann entropy, yielding either a central charge $c=2$ or  $c=1$ [see \ref{sec:appendix_entropy}] in the vortex-liquid phases, corresponding to two- and one-component Luttinger liquids, respectively (we do not distinguish between the $c=1$ and $c=2$ vortex-liquid  phase in the figures).
In principle, the study of current-current correlations could also permit to distinguish the different vortex phases. However, the superposition of the behaviours of the different components 
renders the analysis of correlations less conclusive for finite-size systems.

We stress that the emergence of vortex phases for bosons on the three-leg ladder is a many-body effect: the minimum of the single-particle dispersion
is  always at zero momentum, corresponding to a Meissner phase with counterclockwise chiral-current.
The vortex phases are thus not inherited from finite-momentum global minima in the single-particle dispersion relation, unlike in two-leg ladders~\cite{atala13,huegel14,piraud15,greschner15}.
%

\subsubsection{Staggered-current phase.}
Around $\Jperp \sim J$ and for $\phi \lesssim \pi$, a phase with staggered currents (SC) emerges [see figure~\ref{fig:currentpatterns}(f)].
It is characterized by strong currents going around each plaquette.
This phase breaks translation invariance along the legs of the ladder and has a commensurate vortex density $\rho_v^{\alpha}=1/2$, corresponding to a spatial periodicity of two lattice sites.
It can be visualized as two VL states on the upper and lower two-leg ladder, shifted by one site with respect to each other.
This phase shows a Mott-insulator to superfluid transition when lowering $\Jperp/J$ [see Fig.~\ref{fig:phasediagram} and \ref{sec:appendix_massgap}].
At $\phi=\pi$, the Hamiltonian~(\ref{eq:hamiltonian}) is real, and therefore, time-reversal invariant. The SC phases then spontaneously break this symmetry [see figure~\ref{fig:currentpatterns}(f)]: reversing all currents does not leave the pattern invariant; they can thus be viewed as a generalization of the chiral Mott-insulator and superfluid phases realized on two-leg ladders~\cite{dhar12, dhar13}.

\subsection{Chiral-current reversal in the Meissner phase}
Let us further investigate the region $\phi/\pi \gtrsim 0.75$ where the chirality of the chiral currents is opposed to the one favored by the bare flux. 
In the strong-rung limit $\Jperp \gg J$, the three-site problem is solved analytically and standard perturbation theory (PT) can be applied.
To first order in $J/\Jperp$, the chiral current is then given by
\begin{eqnarray}\label{eq:PT}
 \jC^{PT} = \frac{J^2}{2\sqrt{2}\Jperp} \left( \sin \phi + \frac{3}{4} \sin 2\phi \right),
\end{eqnarray}
which implies a reversal of the current at  $\phi = \arccos (-2/3) \approx 0.73 \pi$, independently of $\Jperp$.
Equation~(\ref{eq:PT}) agrees very well with the DMRG calculations for $\Jperp/J=5$ [see the inset of figure~\ref{fig:meissner}(a)], and the parameter point at which the reversal occurs barely moves for $\Jperp /J \gtrsim 1.6$ [see figure~\ref{fig:phasediagram}].

\begin{figure}[t]
\centering
\includegraphics[width=.7\linewidth]{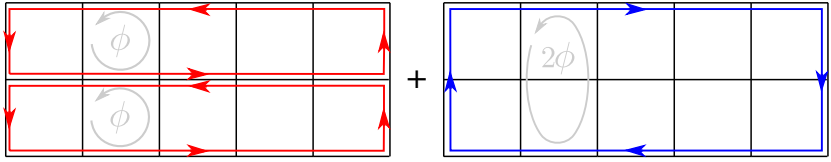}
\caption{Schematic representation of the two contributions to equation~\eref{eq:PT} for $\pi/2<\phi<\pi$.}
\label{fig:sketch-currents}
\end{figure}

The flux dependence $\jC \propto \sin(\phi)$ is directly inherited from that of the two-leg ladder in the Meissner phase~\cite{piraud15}.
The two terms of~\eref{eq:PT} can therefore be interpreted as stemming from two contributions, which are sketched in figure~\ref{fig:sketch-currents}.
The first term (left panel) corresponds to chiral currents flowing on the two sub-ladders (formed by the middle leg and the upper or lower leg) 
resulting in a zero net-current on the middle leg,
whereas the second term (right panel) corresponds to particles propagating only along the top and bottom leg of the three-leg ladder.
This can be thought of as a Meissner phase with an effective flux $\phi_{\rm eff}=2\phi$. 
For $\pi/2 < \phi < \pi$, the second contribution is negative, which can lead to a total chiral current with a reversed chirality.
This reversal is thus associated with a doubling of the effective flux along the rung direction. It can already be captured in the minimal model of two plaquettes ($L=2$), in which a qualitatively similar reversal occurs in the strongly interacting limit.
An increase of the effective flux also underlies the chiral-current reversal in the case of two-leg ladders~\cite{greschner15},
yet there, the increase results from spontaneous breaking of translation symmetry.

%
%

The $\Jperp \gg J$ limit also permits to understand another numerical observation: the density is imbalanced between the legs, in all phases.
PT indeed predicts a density twice as large on the middle leg than on the outer ones. This difference decreases when lowering $\Jperp/J$.

\subsection{Finite interactions $U/J<\infty$}
\begin{figure}[t]
\centering
\includegraphics[width=.7\linewidth]{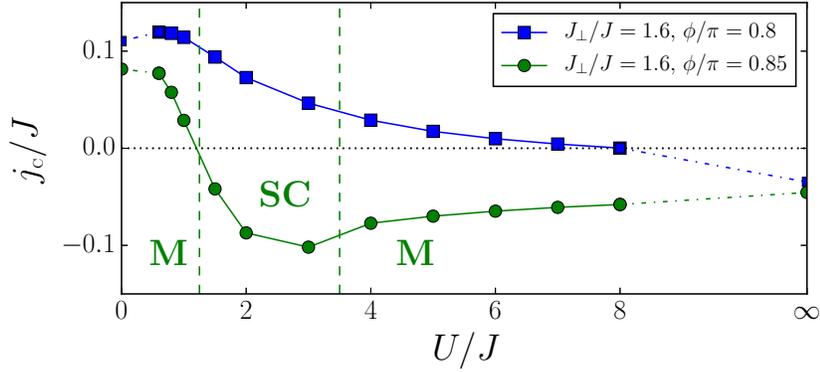}
\caption{Chiral current $\jC$ as a function of on-site interaction $U$ for $\Jperp/J=1.6$ and $\phi/\pi = 0.8$ and $0.85$.
}
\label{fig:meissner_vs_u}
\end{figure}

In the single-particle case ($U/J=0$), the Meissner phase has counterclockwise chirality for any choice of parameters.
Therefore, there must  be a reversal of the chiral current as the interaction strength increases.
This is indeed the case as shown in figure~\ref{fig:meissner_vs_u},  displaying $\jC$ as a function of $U$, for $\phi/\pi=0.8$ and $0.85$ and  $\Jperp/J=1.6$.
The reversal from a negative to a positive current occurs at a finite interaction strength $U$ whose value depends on $\phi$.
For $\phi/\pi=0.85$, the system enters into a staggered-current phase at intermediate values of $U$, as indicated in the figure.
Translational invariance is spontaneously broken in the SC phase and the unit cell  comprises four plaquettes [see figure~\ref{fig:currentpatterns}(f)].
For $\phi \in [3\pi/4,\pi]$, the effective flux is $\phi_{\rm eff}=4\phi \in [-\pi,0]$ modulo $2\pi$, and the current is reversed.
This realizes another instance of the chiral-current reversal due to spontaneously enlarged unit cells first presented in reference~\cite{greschner15}.
The presence of the SC phases stabilizes the current reversal down to much smaller values of $U/J$ compared to parameters for which the SC phase is absent 
(compare the data for $\phi/\pi=0.8$ and $0.85$ shown in figure~\ref{fig:meissner_vs_u}).
Note that the VL phase leads to the same unit-cell enlargement.
However, as it occurs for $\phi \in [\pi/2,3\pi/4]$ in the HCBs case (see figure~\ref{fig:phasediagram}), one has $\phi_{\rm eff}=4\phi \in [0,\pi]$ modulo $2\pi$, and the chiral current is not reversed.

%
%

\section{Experimental realizations}
\begin{figure}[t]
\centering
\includegraphics[width=.7\linewidth]{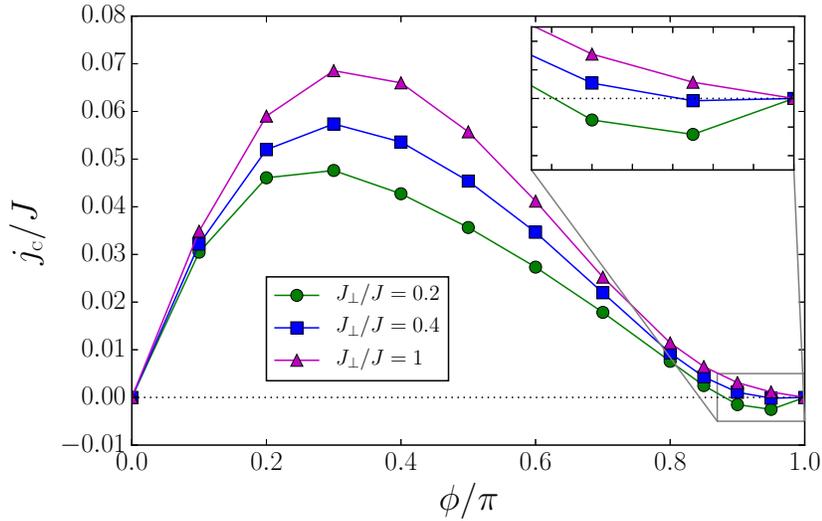}
\caption{Chiral current $\jC$ [cf.~equation~\ref{eq:meissner}] as a function of the magnetic flux $\phi$ for $\Jperp/J=0.2$, 0.4 and $1$, for long-range interactions along the rungs [cf.~equation~\ref{eq:LRint}] and $U/J=10$.
}
\label{fig:longrangeint}
\end{figure}

The most straightforward experimental approach would be to use a superlattice to split a two-dimensional lattice into three-leg ladders~\cite{foelling07}.
This would create an energy offset between the middle and the outer two legs, which could easily be compensated by another superlattice with a two-site periodicity.
For systems with a synthetic lattice dimension~\cite{stuhl15,mancini15,celi14}, the interaction in the rung direction is not on-site as considered here,
but depends on the total density $n^{\rm t}_{r}=\sum_{\ell=1}^{3} n_{r,\ell}$ in all three rung sites. Thus, the appropriate form is
\begin{equation}
H_{\rm int}=\frac{U}{2} \sum_{r=1}^{L} n^{\rm t}_{r}(n^{\rm t}_{r}-1). 
\label{eq:LRint}
\end{equation}
Our DMRG results for $U/J=10$ show that this type of long-range interactions suppresses vortex phases.
This is qualitatively consistent with the study of the two-leg ladder, in which this type of interaction also favors the Meissner phase over the vortex phase~\cite{petrescu13,petrescu15}.
The chiral-current reversal in the Meissner phase at a filling of one-third survives at small $\Jperp/J \lesssim 0.3$ [see figure~\ref{fig:longrangeint}].

%

\section{Summary}
We presented the phase diagram of strongly-interacting bosons on a three-leg ladder  subjected to a homogeneous magnetic flux. We identified 
several phases, including vortex-liquid phases, a vortex lattice and a Meissner phase.
Moreover, there is a state with staggered currents, which leads to a reversal of the chiral current due to the spontaneous increase of the unit cell, similar to the 
situation discussed in~\cite{greschner15}.
Fascinatingly, the Meissner phase also shows a chiral-current reversal when increasing the strength of the magnetic flux per plaquette, yet 
in the Meissner phase, translational invariance is not broken. 
The analysis of the strong-rung and large-interaction limit indicates that a doubled flux is experienced along the rung direction.
Therefore, the chiral-current reversal in the Meissner phase is qualitatively different.
We argue that interactions are a necessary ingredient to obtain these behaviors.

\ack
We are grateful to V. Alba, T. Giamarchi, S. Greschner, A. Honecker, I. Spielman, and T. Vekua for helpful discussions.
U.S. acknowledges funding by DFG through NIM and SFB/TR 12. I.P.M. acknowledges the support from the Australian Research Council Centre of Excellence for Engineered Quantum Systems, CE110001013, and the Future Fellowships scheme, FT100100515. 
The research of M.P. was supported by the European Union through the Marie-Curie grant ToPOL (No. 624033) (funded within FP7-MC-IEF). F.H.-M. and M.P. also acknowledge support from the Institute for Nuclear Theory during the program INT-15-1 "Frontiers in Quantum Simulation with Cold Atoms".

\appendix
\section{Entanglement entropy}\label{sec:appendix_entropy}
The entanglement entropy $S_{vN}$ is defined as the von Neumann entropy corresponding to a bipartition of the wavefunction into two subsystems $A$ and $B$:
\begin{eqnarray}
 S_{vN} = -\Tr \rho_{A}\ln \rho_{A},
\end{eqnarray}
where $\rho_{A} = \Tr_{B} |\psi\rangle \langle \psi|$ is the reduced density matrix of subsystem $A$. 
In conformal field theory, for a one-dimensional system with open boundary conditions and total length $L$, the von Neumann entropy of a subsystem of length $r$ is given by~\cite{Vidal2003a, Calabrese2004}
\begin{eqnarray}\label{eq:svn}
S_{vN}(r) = \frac{c}{6}\log\left[\frac{2L}{\pi}\sin\left(\frac{\pi r}{L}\right) \right] + g,
\end{eqnarray}
where $c$ is the central charge of the conformal field theory and $g$ is a non-universal constant. The central charge measures the number of gapless modes in the system.

\begin{figure}[ht]
\centering
\includegraphics[width=1.\linewidth]{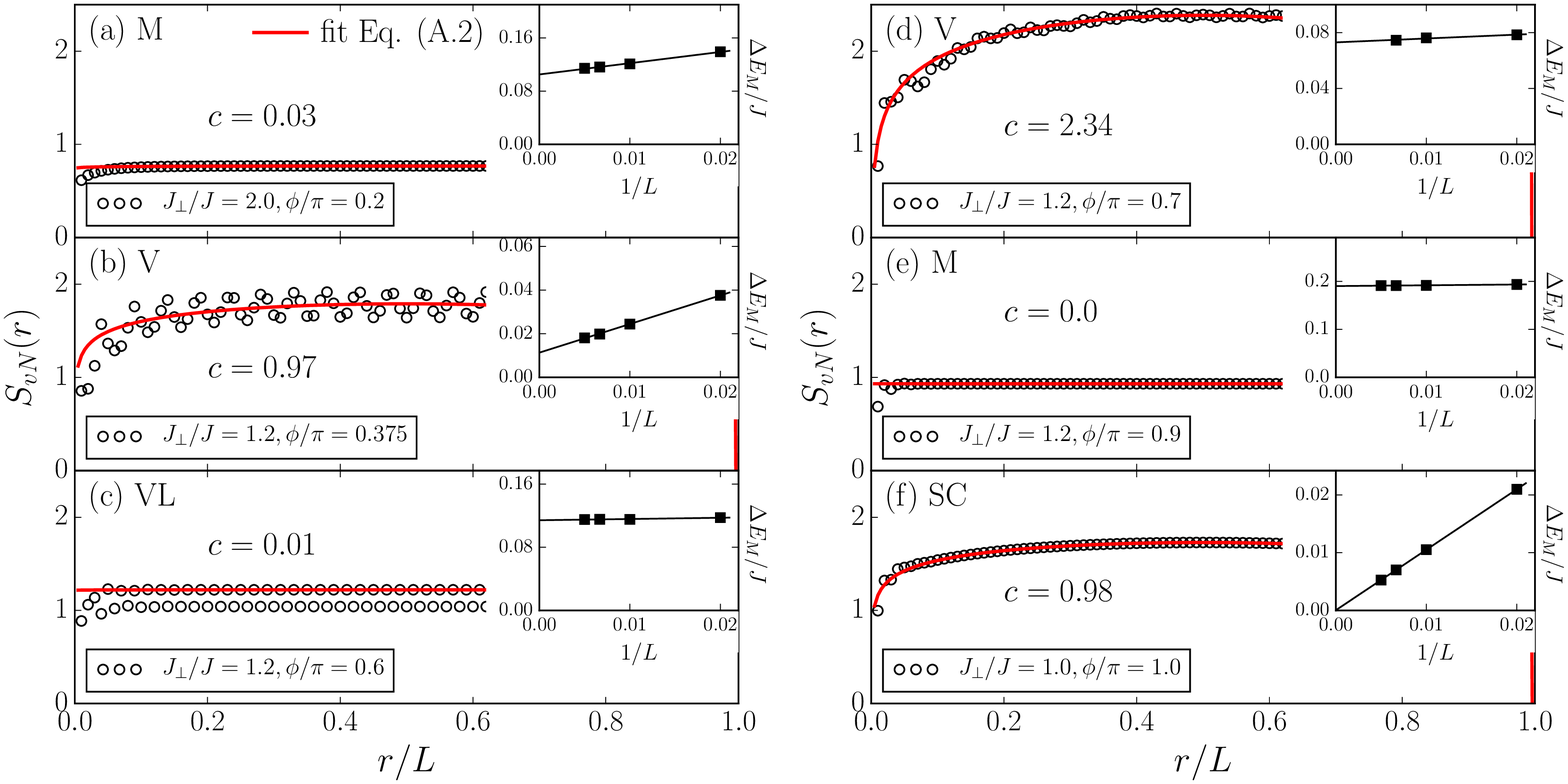}
\caption{
 The entanglement entropy $S_{vN}$ as a function of the relative length of the subsystem $r/L$ for $L=100$. Symbols show DMRG results for (a) $\Jperp/J=2.0$, $\phi/\pi=0.2$ in the Meissner phase; (b) $\Jperp/J=1.2$, $\phi/\pi=0.375$ in the vortex-liquid phase; (c) $\Jperp/J=1.2$, $\phi/\pi=0.6$ in the vortex-lattice phase; (d) $\Jperp/J=1.2$, $\phi/\pi=0.7$ in the vortex-liquid phase; (e) $\Jperp/J=1.2$, $\phi/\pi=0.9$ in the Meissner phase with currents with reversed chirality; (f) $\Jperp/J=1.0$, $\phi/\pi=1.0$ in the staggered-current phase. The red line is a fit of~\eref{eq:svn} to the DMRG data. The insets show the mass gap $\Delta E_{M}$ as a function of inverse system length which is extrapolated to the infinite length limit with a linear fit.
}
\label{fig:svn}
\end{figure}

In figure~\ref{fig:svn} we show the entanglement entropy $S_{vN}(r)$ for straight cuts through the three-leg ladder between rungs $r$ and $r+1$. By fitting~\eref{eq:svn} to the data we obtain estimates for the central charge. Oscillations of the entropy in the vortex phases make the fit very sensitive to the fitting region used. This applies, in particular, to the vortex-liquid phases. Nevertheless, the behavior of the central charge provides further insights into the nature of the commensurate-incommensurate transition.
%

In the Meissner phase [figures~\ref{fig:svn}(a) and (e)] the values of the central charge are very close to zero in agreement with an entanglement entropy independent of the subsystem size, which corresponds to the area law of a fully gapped phase~\cite{Eisert2010}.

The vortex phases can be divided into vortex-liquid and vortex-lattice phases with incommensurate and commensurate vortex densities, respectively. As illustrated in figure~{\color{red}5} of the main text there are two vortex densities that undergo a transition from the Meissner to a vortex-liquid and vortex-lattice phase with increasing flux $\phi/\pi$ and at low enough $\Jperp/J$. This picture is consistent with the behavior of the central charge. More precisely, at $\Jperp/J=1.2$ and $\phi/\pi=0.375$, we find one incommensurate vortex density in the Fourier transform of the rung currents and a central charge of $c\approx 1$ [figure~\ref{fig:svn}(b)]. At flux $\phi/\pi=0.7$, there are two incommensurate vortex densities. The central charge is $c\approx 2$ in agreement with the presence of two gapless modes [figure~\ref{fig:svn}(d)]. At $\phi/\pi=0.6$, both vortex densities are commensurate and we find an entanglement entropy independent of subsystem size (up to oscillations due to the increased unit cell), or $c=0$ in terms of the central charge [figure~\ref{fig:svn}(c)].

This suggests that the modes, which are measured by the central charge, can be related to the dominating vortex densities in the Fourier transform of the rung currents. An incommensurate vortex density would be associated with a gapless mode while a commensurate vortex density would correspond to a gapped mode.

In the staggered-current phase the vortex densities are commensurate, while the fit gives a central charge $c \approx 1$.
But the scaling of the entanglement spectrum indicates that it is two independent $c=0.5$ modes, and also the scaling of the correlation function gives the scaling dimension $\Delta \simeq 1/8$ (not shown), consistent with $c=0.5$.
This indicates that the mass gap is closed, which is compatible with the analysis of the mass gap and the Binder cumulant presented in the next section.

\section{Mott insulator to superfluid transition}\label{sec:appendix_massgap}
The mass gap $\Delta E_{M}$ is defined as
\begin{eqnarray}
 \Delta E_{M} = \frac{1}{2} \left[ E(N-1) + E(N+1) \right] - E(N),
\end{eqnarray}
where $E(N)$ is the total ground-state energy of the system with $N$ particles.

We generally observe a finite mass gap that becomes smaller when decreasing the inter-leg coupling $\Jperp$. In the $\Jperp=0$ limit, the mass gap is known to be closed. As already mentioned in the main text, in the $\phi/\pi=0$ limit, the transition to the superfluid phase at finite $\Jperp/J$ can be expected to be of the Berezinskii-Kosterlitz-Thouless type~\cite{cabra97,cabra98}, and, therefore, is very hard to pinpoint numerically. 

In the insets of figure~\ref{fig:svn} we show the mass gap as a function of the inverse system size $1/L$ extrapolated with a linear fit to the thermodynamic limit. We find a finite mass gap in the Meissner phase for $J/\Jperp=2$, $\phi/\pi=0.2$  and $J/\Jperp=1.2$, $\phi/\pi=0.9$ [figures~\ref{fig:svn}(a) and (e)].
The mass gap closes in the Meissner phase when lowering $J/\Jperp$. The vortex-liquid phases [figures~\ref{fig:svn}(b) and (d)] and the vortex-lattice phase [figure~\ref{fig:svn}(c)] have finite mass gaps with weak system-size dependence. The finite mass gap in the vortex-liquid phases confirms that the charge mode does not contribute to the values of the central charges in figures~\ref{fig:svn}(b) and (d) but they are fully accounted for by the two other modes on the three-leg ladder that undergo the commensurate-incommensurate transition.

The analysis of the central charge is more difficult in the regions where the mass gap can not be resolved (or is actually zero) with the system sizes studied in this work. This is the case in the Meissner phase for $\Jperp/J \lesssim 1.6$ and $\phi/\pi \lesssim 0.3$
and in the staggered-current phase [see the inset of figure~\ref{fig:svn}(f)].

Calculating energy gaps is a quite difficult calculation when the gap is small, so where an order parameter exists it is generally better to use it instead to detect the phase boundary of the Mott-insulating region.
A suitable order parameter for a Mott phase is the non-local string order 
parameter~\cite{emanuele06}, given by
\begin{equation}
O_p^2 = \lim_{|j-i| \rightarrow \infty} \left\langle \Pi_{r=i}^j (-1)^{1-n^{\rm t}_{r}} \right\rangle \; ,
\end{equation}
where $n^{\rm t}_{r} = \sum_{\ell =1}^{3} n_{r, \ell}$ is the total particle density on rung $r$.

This can be expressed in terms of a `kink' operator,
\begin{equation}
p(i) = \Pi_{r < i} (-1)^{1-n^{\rm t}_{r}} \; ,
\end{equation}
from which $O_p^2$ takes the form of a correlation function,
\begin{equation}
O_p^2 = \lim_{|j-i| \rightarrow \infty} \left\langle p(i)p(j) \right\rangle \; ,
\end{equation}
or, we can construct an extensive order parameter $P = \sum_i p(i)$. We cannot
evaluate $\langle P \rangle$ directly, since the sign is indeterminate, however, we
can calculate $\langle P^2 \rangle$, which is simply related to $O_p^2$, as
\begin{equation}
O_p^2 = \langle P^2 \rangle / L^2 \; ,
\end{equation}
where $L$ is the length of the lattice.
The great advantage of this construction is that the $P$ operator is easy to construct
for matrix-product-state (MPS) algorithms using the triangular matrix-product-operator (MPO) formulation~\cite{mcculloch07,mcculloch08},
and the expectation value of such triangular MPO's can be evaluated directly in the
thermodynamic limit~\cite{michel10}. This allows us to calculate the Binder cumulant~\cite{binder81}
of $P$,
\begin{equation}
B = 1 - \frac{\langle P^4 \rangle}{3\langle P^2 \rangle^2} \; .
\end{equation}
For an infinite matrix-product-state (iMPS), the expectation value of the $n$-th power of an operator $P^n$ is obtained
as a degree $n$ polynomial of the lattice size $L$, which is exact in the asymptotic large-$L$
limit. Hence the quantity $O_p^2 = \langle P^2 \rangle / L^2$ 
is evaluated directly as the coefficient of
the degree 2 component of $\langle P^2 \rangle$.

If we evaluate the Binder cumulant directly for an iMPS in the large $L$ limit,
then it simply probes where $O_p^2$ is zero or non-zero, so does not give any
additional information. 
Since the coefficient of the $n$-th degree term in $\langle P^n \rangle$
is simply $O_p^n$, it follows that if $O_p^2$ is non-zero then the Binder cumulant in the
large $L$ limit is identically
equal to $2/3$. Similarly, if $O_p^2 = 0$, then the cumulant expansion of
$\langle P^4 \rangle$ reduces to $3\langle P^2 \rangle^2$ and the Binder cumulant is identically
zero. Hence we have a step function located at the point where $O_p^2$ becomes non-zero, which
is what one would expect if one takes the large-$L$ limit of a finite-size Binder cumulant.

\begin{figure}
\centering
\includegraphics[width=1\textwidth]{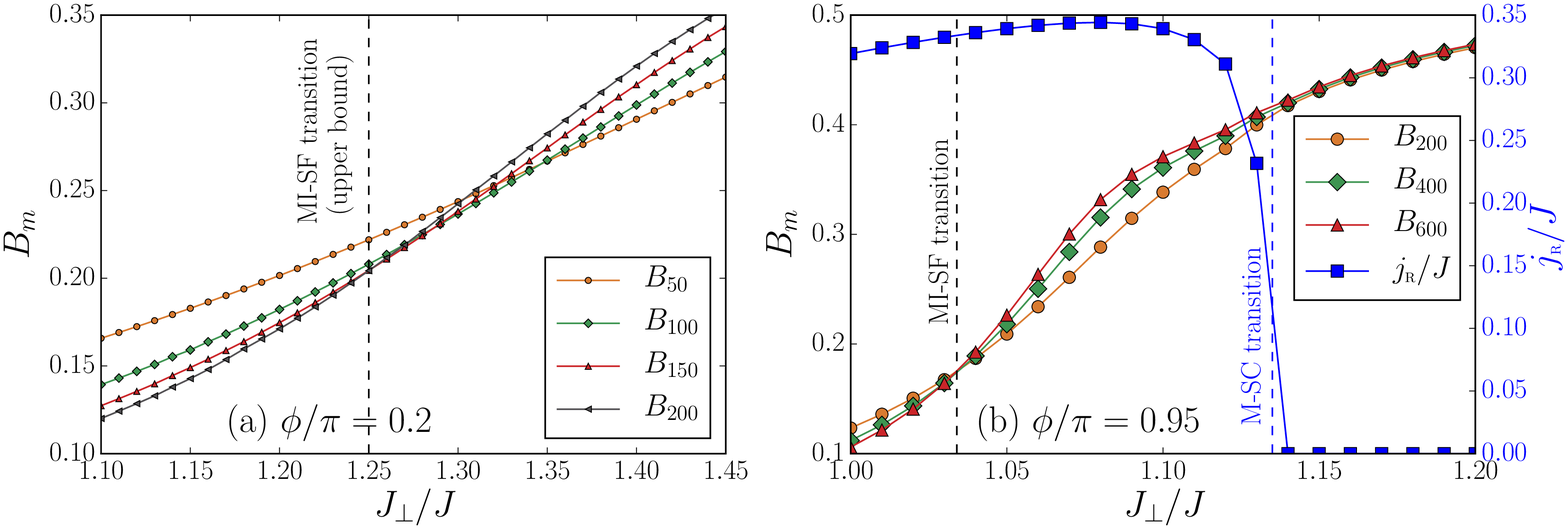}
\caption{Binder cumulant $B_m$ for various number of states $m$, for (a) $\phi/\pi=0.2$ and (b) $\phi/\pi=0.95$, to analyze the Mott insulator to superfluid transition.
In (a) the crossing point moves towards lower $\Jperp/J$ as $m$ increases, compatible with a Berezinskii-Kosterlitz-Thouless transition at $\Jperp/J < 1.25$.
In (b) the crossing point does not move significatively when $m$ increases, compatible with a second-order transition at $\Jperp/J \simeq 1.035$.
The average rung current $\jR$, as defined in the main text, is also displayed. It shows that the Meissner to SC transition occurs at $\Jperp/J \simeq 1.135$.}
\label{fig:binder}
\end{figure}
However, for an iMPS the correlation length is always finite, so one might expect that
finite-entanglement-scaling with the basis size $m$ can be used instead of finite-size scaling. 
This is indeed the case,
if we evaluate the polynomial $\langle P^n \rangle$ at $L = b \xi$, where $\xi$ is the correlation
length, and $b$ is some scaling factor. This is equivalent to calculating the order parameter
over a finite section of size $L = b \xi$ of the infinite lattice.

Since $\xi$ depends on the number of states $m$,
we can plot a family of curves of the Binder cumulant
$B_m$ for different values of $m$.
For a finite system, the value of the Binder cumulant at a second-order critical point is independent of $L$
(up to higher order corrections). For an iMPS, the analogous result is that the
critical value of the Binder cumulant is independent of $m$.
Note that for finite size systems the actual value of the Binder cumulant 
at the critical point is
not universal and depends on the boundary conditions~\cite{selke06}. For an iMPS, the value
of the $m$-dependent Binder cumulant depends on the chosen scale factor $b$. For the
calculations here, we use $b=1$.

As an example, we show in~\fref{fig:binder}(a) the
Binder cumulant as a function of $\Jperp$ for $\phi/\pi=0.2$.
The data is compatible with a Berezinskii-Kosterlitz-Thouless transition from the Meissner Mott-insulating region into the Meissner superfluid region as $\Jperp/J$ decreases and permits to give an upper bound for the critical point, which is reported in \fref{fig:phasediagram} in the main text.
For  $\phi/\pi=0.95$ [see~\fref{fig:binder}(a)], a Mott-insulator to superfluid transition also occurs in the SC phase and it is compatible with a second-order critical point at $\Jperp/J \simeq 1.035$.
The Meissner to SC phase transition, which can be characterized by a rise of the average rung-current $\jR$, occurs at a distinct point $\Jperp/J \simeq 1.135$.

\section*{References}
\bibliographystyle{iopart-num}
\providecommand{\newblock}{}

\end{document}